\documentclass[twocolumn,showpacs,preprintnumbers,amsmath,amssymb,prd,amsfonts,floats,superscriptaddress,floatfix,nofootinbib]{revtex4}

\usepackage{graphicx}
\usepackage{dcolumn}
\usepackage{bm}
\usepackage{longtable}
\usepackage{epstopdf}

\begin{document}
\bibliographystyle{apsrev}

\title{$\Lambda\alpha$DM:  Observational constraints  on   unified dark matter \\with constant speed of sound}

\author{Amedeo Balbi}
\affiliation{Dipartimento di Fisica, Universit\`a di Roma ``Tor Vergata'',
via della Ricerca Scientifica 1, 00133 Roma, Italy}
\affiliation{INFN Sezione di Roma ``Tor Vergata'',
via della Ricerca Scientifica 1, 00133 Roma, Italy}
\author{Marco Bruni}
\affiliation{Institute of Cosmology and Gravitation, University
of Portsmouth, Mercantile House, Portsmouth PO1 2EG, United Kingdom}
\affiliation{Dipartimento di Fisica, Universit\`a di Roma ``Tor Vergata'',
via della Ricerca Scientifica 1, 00133 Roma, Italy}
\affiliation{INFN Sezione di Roma ``Tor Vergata'',
via della Ricerca Scientifica 1, 00133 Roma, Italy}
\author{Claudia Quercellini}
\affiliation{Dipartimento di Fisica, Universit\`a di Roma ``Tor Vergata'',
via della Ricerca Scientifica 1, 00133 Roma, Italy}

\begin{abstract}
We consider the hypothesis that dark energy  and dark matter are the two faces of a single  dark component, a unified dark matter (UDM) that we assume can be modeled by  the affine equation of state (EoS)   $P= p_0 +\alpha \rho$, resulting in an {\it  effective cosmological constant} $\rho_\Lambda=-p_0/(1+\alpha)$. The affine EoS arises from the simple assumption that the speed of sound is constant; it may be seen as an approximation to an unknown barotropic EoS $P=P(\rho)$, and may as well represent the tracking solution for the dynamics of a scalar field with  appropriate potential.
Furthermore, in principle the affine EoS  allows the UDM to  be phantom. We constrain the parameters of the model, $\alpha$ and $\Omega_\Lambda$, using 
 data from a suite of different cosmological observations, and perform a  comparison with the standard $\Lambda$CDM model,
containing both cold dark matter and a cosmological constant.
First considering a flat  cosmology, we find that the UDM   model with affine EoS fits the joint observations very well, better than   $\Lambda$CDM, with best fit values $\alpha=0.01 \pm 0.02$ and $\Omega_\Lambda=0.70 \pm 0.04$ (95\% confidence intervals). The standard model (best fit $\Omega_\Lambda=0.71\pm 0.04$), having one less parameter, is  preferred by a Bayesian model comparison. However, the affine EoS is at least as good as  the standard model  if a flat curvature  is not assumed as a prior for $\Lambda$CDM.
 For the latter, the best fit values are $\Omega_K=-0.02^{+0.01}_{-0.02} $ and $\Omega_\Lambda=0.71 \pm 0.04$, i.e. a closed model is preferred.
A phantom UDM  with affine EoS is ruled out well beyond $3\sigma$.
\end{abstract}


\maketitle

\section{Introduction}
In the past few years, evidence that the universe is going through a phase of accelerated expansion has become compelling~\cite{Riess:astro-ph/9805201,Perlmutter:astro-ph/9812133,Riess:astro-ph/0402512,Riess:astro-ph/0611572}. The cause for the acceleration, however, remains mysterious, and might be regarded as the most outstanding problem in contemporary cosmology. The favoured working hypothesis is to consider a dynamical, almost homogeneous component with negative pressure,  dubbed  dark energy \cite{Peebles:1987ek,Wetterich:1987fm,Caldwell:astro-ph/9708069,Wang:astro-ph/9901388}. Such a framework helps alleviating a number of fundamental problems arising when an {\it ad hoc}  cosmological constant $\Lambda$ term in Einstein equations is interpreted as the energy density of the vacuum~\cite{Weinberg:1988cp,Peebles:astro-ph/0207347,Padmanabhan:hep-th/0212290}.
Alternatives include modifications of gravity and extra dimensions~\cite{Sahni:astro-ph/0502032,Maartens:astro-ph/0602415,Copeland:hep-th/0603057}.

Here, within  the framework of General Relativity (GR),  we explore the  hypotesis that dark energy (the source for the observed acceleration of the universe) and dark matter (DM, required to explain structure formation) are the two faces of a single  dark component, a unified dark matter (UDM) that we assume can be modeled by  a simple 2-parameter barotropic equation of state (EoS). In turn, this EoS can be derived from the simple assumption that the UDM speed of sound is constant.

  Recent analyses of cosmological data suggest that within the  standard $\Lambda$CDM scenario (Cold DM plus $\Lambda$) a slightly closed, positively curved model is preferred \cite{Spergel:astro-ph/0603449}. With this in mind, motivated by a  theoretical bias in favor of a flat inflationary scenario, 
  we   compare our UDM model, assumed to be flat, with both   flat and curved $\Lambda$CDM models, using Bayesian methods \cite{Liddle:astro-ph/0401198,Liddle:astro-ph/0608184}.

\section{The Model} \label{model}
Let's consider a flat Friedmann-Robertson-Walker universe in GR, with radiation, baryons and a single UDM component with energy density $\rho_X$. The dynamics of this model is governed by the Friedmann equation
\begin{equation}
\label{einstein1}
H^2=\frac{8\pi G}{3}(\rho_r+\rho_b+\rho_X)
\end{equation}
and the energy  conservation equations  of the three components. Here $H$ is the Hubble expansion scalar related to the scale factor $a$ by $H=\dot{a}/a$, and no  {\it ad hoc} cosmological constant $\Lambda$ term is assumed in Eq. (\ref{einstein1}). 

With the usual scaling laws $\rho_b\propto a^{-3}$ and $\rho_r\propto a^{-4}$ for the density of  baryons and radiation, we now assume that the dark component is represented by a barotropic fluid with EoS $P_X=P_X(\rho_X)$, satisfying  the  conservation equation
\begin{equation}
\label{einstein2}
\dot{\rho}_X=-3H(\rho_X+P_X).
\end{equation}
It is clear from Eq.\ (\ref{einstein2}) that, if  there exists  an energy density value  $\rho_X=\rho_\Lambda$ such that $P_X(\rho_\Lambda)=-\rho_\Lambda$, then $\rho_\Lambda$ has the dynamical role of an {\it effective} cosmological constant: $\dot{\rho}_\Lambda=0$ (see \cite{Ananda:astro-ph/0512224} for a detailed discussion).

In order to provide for acceleration, our UDM component must violate the strong energy condition (SEC) (see e.g.\ \cite{Visser:1997aa}): $P_X<-\rho_X/3$ at least below some redshift. This can be achieved by a constant $w_X=P_X/\rho_X$,  which is indeed allowed by observational  tests  based  only  on  the  homogeneous  isotropic  background  evolution, see e.g.\   \cite{Davis:astro-ph/0701510,Maartens:astro-ph/0603353} (tests of the  same  kind  we  are  going  to  consider here, see below).
In this case, however,  one  would  have $c_s^2=dP_X/d\rho_X=w<0$ for the adiabatic speed of sound, and  this would have nasty consequences for structure formation in an adiabatic fluid scenario\footnote{The EoS with constant $w$ may as well represent a scalar field with exponential potential \cite{LM85}, and this would have a unitary effective speed of sound, $c^2_{eff}=1$, see e.g.\ \cite{BED91,Hu:astro-ph/9801234} and references therein. }. Instead, 
given that the EoS $P_X=P_X(\rho_X)$ is unknown, the next  simplest approximation we can make to model it is to assume a constant speed of sound  $dP_X/d\rho_X\simeq \alpha$, leading to the  2-parameter affine form \cite{Ananda:astro-ph/0512224}
\begin{equation}
\label{EOSaffine}
P_X\simeq p_0+\alpha \rho_X.
\end{equation}
This allows for violation of SEC even with \mbox{$c_s^2=\alpha \geq 0$}. Then, using (\ref{EOSaffine}) in (\ref{einstein2})  and asking for $\dot{\rho}_\Lambda=0$ leads to the {\it effective cosmological constant } $\rho_\Lambda=-p_0/(1+\alpha)$.
Eq.\ (\ref{EOSaffine}) may also be regarded (after regrouping of terms) as the Taylor expansion, up to ${\cal O}(2)$, of {\it any} EoS $P_X=P_X(\rho_X)$ about the present energy density value $\rho_{Xo}$ \cite{Visser:gr-qc/0309109}.
In addition, it may also represent the exact tracking solution for the dynamics of a scalar field with appropriate potential \cite{Quercellini:2007ht}.

The EoS (\ref{EOSaffine}), if taken as an approximation, could be used to parametrize a dark component (either UDM or DE) at low and intermediate redshift. In the following we are going to make a more radical assumption, that is   we are going to extrapolate the validity of Eq.\ (\ref{EOSaffine}) to any time. In doing this  we are therefore going to build a cosmological model based on a  UDM component with EoS (\ref{EOSaffine}), to be tested against observables at low and high redshift, as described in the next section. 
 
 Using the EoS (\ref{EOSaffine}) in the conservation equation \ (\ref{einstein2}) leads to a simple evolution for $\rho_X(a)$:
\begin{equation}
\label{rhodia}
\rho_X(a)=\rho_\Lambda+(\rho_{Xo}-\rho_\Lambda)a^{-3(1+\alpha)},
\end{equation}
where today $a_o=1$. Formally, with the EoS (\ref{EOSaffine}) we can then interpret our UDM   as made up of the {\it effective} cosmological constant $\rho_\Lambda$ and an evolving part with present ``density" $\tilde{\rho}_m=\rho_{Xo}-\rho_\Lambda$. We may then dub it $\Lambda\alpha$DM\footnote{It should be clear that, for $\tilde{\rho}_m>0$ and starting from $c_s^2=\alpha={\rm const.}$, our model is formally totally equivalent to a model where a standard cosmological constant $\Lambda=8\pi G\rho_\Lambda$  is assumed {\it a-priori} in Einstein equations, i.e.\  in (\ref{einstein1})), together with a standard fluid with linear EoS $w=\alpha$ and present density $\tilde{\rho}_m$. Indeed, by definition $\dot{\rho}_\Lambda=0$,  thus it is easy to check that \ $\rho_\Lambda$  and $\rho_m \equiv \rho_X-\rho_\Lambda$ in (\ref{rhodia}) separately satisfy  their  own conservation equations, so that there is no coupling between the dark energy part $\rho_\Lambda$ and the dark matter part $\rho_m$ of our UDM fluid.}.
{\it A priori}, no restriction on the values of $\alpha$ and $p_o$ is required, but  one needs $p_o<0$ and $\alpha>-1$ in order to satisfy the conditions that $\rho_\Lambda>0$ and   $\rho \rightarrow \rho_\Lambda$ in the future, i.e. to have that $\rho_\Lambda$ is an attractor for Eq.\ (\ref{EOSaffine}). Then, our UDM is phantom 
if $\tilde{\rho}_m <0$, but without a ``big rip", cf.\ \cite{Ananda:astro-ph/0512224}. Last, but not least, it follows from Eq.\ (\ref{rhodia}) that, with $\tilde{\rho}_m>0$ and $\alpha=0$, our model is  equivalent to a $\Lambda$CDM. This allows for a straightforward comparison of models, done in section \ref{results}.

Finally, the EoS of our $\Lambda\alpha$DM model can be characterized by its parameter $w_X=P_X/\rho_X$:
\begin{eqnarray}
w_X & = & -(1+\alpha)\frac{\rho_\Lambda}{\rho_X}+\alpha, \\
w_{Xo}  & \simeq&  -\Omega_\Lambda +\alpha(1-\Omega_\Lambda) -\Omega_\Lambda(1-\Omega_{Xo}) +\cdots \label{eq:w0}
\end{eqnarray}
where today's value $w_{Xo} $ is approximated using $\Omega_{Xo}\approx 1$ and assuming $|\alpha|\ll1$.
 Hence at leading order $w_{Xo} \simeq -\Omega_\Lambda $, while $w_X\rightarrow -1$ asymptotically in the future.

 \section{Observables}\label{observables}
In order to constrain the affine EoS (\ref{EOSaffine}) we perform maximum likelihood tests on its parameters, using different cosmological probes. We assume a flat cosmological model, $\Omega=1$, and we hold the baryon density fixed to the best fit value derived from the analysis of WMAP 3 year data $\Omega_b h^2=0.02229\pm 0.00075$~\cite{Spergel:astro-ph/0603449}. We also  fix the   Hubble parameter to the value measured by the Hubble Space Telescope (HST) Key Project $H_0=72\pm 8$ km/s/Mpc \cite{Freedman:astro-ph/0012376} (but see below about marginalization). This leaves only two free parameters in our analysis, i.e.\ those  characterizing the affine EoS: the constant speed of sound  $c_s^2=\alpha$ and $\Omega_\Lambda=8\pi G\rho_\Lambda/(3H_o^2)$. Given that our model contains an effective cosmological constant parametrized by $\Omega_\Lambda$ and  that for $\tilde{\rho}_m>0$ in (\ref{rhodia}) and $c_s^2=\alpha=0$ this model is equivalent to a $\Lambda$CDM, in essence we are going to contrain with observables the possible variations of the speed of sound of the overall dark component from the value it takes in the $\Lambda$CDM case, under the simple assumption that $c_s^2=\alpha$ is a constant.

The observables we use have become the standard tools to probe the background cosmology, through their dependence on the expansion rate of the universe (see, e.g. \cite{Davis:astro-ph/0701510, Maartens:astro-ph/0603353}). 
They are:

\paragraph{The present age of the universe, $t_0$.} This is   obtained directly from the definition of $H$.
We compared the theoretical prediction on $t_0$ with its best estimate $t_0=12.6^{+3.4}_{-2.4}$ Gyr, derived from a combination of different astrophysical probes in~\cite{2003Sci...299...65K}. 

\paragraph{The luminosity distance of type Ia supernovae.}
Type Ia SNe light curves allow a determination of an extinction-corrected distance moduli,
\begin{equation}
\mu_0=m-M=5\log \left(d_L/{\rm Mpc}\right)+25\label{distance_modulus}
\end{equation}
where $d_L=(L /4\pi F)^{1/2}=(1+z)\int_0^z dz' / H(z')$ is the luminosity distance. We compare our theoretical predictions to the values of $\mu_0$ estimated in \cite{Riess:astro-ph/0611572} using a sample of 182 type Ia SNe (the new Gold dataset  \cite{Riess:astro-ph/0611572} containing the old Gold sample \cite{Riess:astro-ph/0402512}, the Supernovae Legacy Survey (SNLS) \cite{Astier:astro-ph/0510447}  and 12 new SNe observed by the HST).

\paragraph{The location of cosmic microwave background (CMB) acoustic peaks.} When the geometry of the universe is held fixed at $\Omega=1$, this only depends on the amount of  dark matter, through the shift parameter $\cal R$ defined as:
\begin{equation}
{\cal R}=\Omega^{1/2}_mH_0 D_A(z_{ls}),
~~~
D_A(z_{ls})=\int_0^{z_{ls}} \! \! \!  \frac{dz'}{H(z')},
\end{equation}
where $D_A$ is the distance to the last scattering surface. We have identified $\rho_m=\tilde{\rho}_m$ for our model, i.e.\ $\Omega_m=8\pi G\tilde{\rho}_m/(3H_o^2)$, and the redshift of last scattering has been estimated using the fit function from \cite{2001ApJ...549..669H}, which for our choice of $\Omega_b h^2$ gives $z_{ls}=1089$.
We derived a value of ${\cal R}$ using five Monte Carlo Markov chains produced in the most recent analysis of the WMAP data\footnote{See  the Lambda web site: http://lambda.gsfc.nasa.gov.}. We estimated a value ${\cal R}=1.71\pm 0.03$. Our new analysis updates previously published values of ${\cal R}$, e.g.\ \cite{Wang:astro-ph/0604051}.

\paragraph{Baryon acoustic oscillations.} The recent detection of acoustic features in the matter power spectrum measured with the Sloan Digital Sky Survey (SDSS) galaxy survey constrains the $A$ parameter defined as:
\begin{equation}\label{eq:AA}
A=\frac{\Omega^{1/2}_mH_0}{0.35\, c} D_V,
~~~
D_V=\left[D_A^2(z)\frac{cz}{H(z)}\right]^{\frac{1}{3}}_{z=0.35}
\end{equation}
where $D_V$ is the distance to $z=0.35$ taking into account the distortion along the line of sight due to the redshift. The value of the $A$ parameter adopted in our analysis is the one measured from the SDSS luminous red galaxy survey: $A=0.469\pm 0.017$ (for $n_s=0.98$) \cite{Eisenstein:astro-ph/0501171}.

We perform our likelihood analysis by sampling the parameter space in the range $[-0.3,0.3]$ for $\alpha$ (where the upper limit is due to the requirement that the UDM fluid does not scale faster than radiation) and $[-0.3,0.9]$ for $\Omega_\Lambda$. The size of our parameter space does not pose any serious problem in terms of computational time. Thus, we preferred to perform a direct evaluation of the likelihood on a predefined grid in the volume under consideration, rather than adopting the now popular Monte Carlo Markov chain approach. When calculating confidence interval in the $\alpha-\Omega_\Lambda$ plane from the SNe data, we always marginalize over a calibration uncertainty (treated as a nuisance parameter) on the absolute magnitude. 
Note that marginalizing over  calibration uncertainty is completely equivalent to  marginalizing  over $H_0$, since both parameters effectively act as an additional term in the distance modulus  (\ref{distance_modulus}).

A final caveat on the use of the $R$ and $A$ parameters. The numerical values adopted in our analysis were both derived assuming an underlying $\Lambda$CDM cosmology. In principle, one might question whether it is a good assumption to adopt the same numerical values when constraining a different class of models. We believe that this is a reasonable approximation when comparing the evolution of the expansion rate with observations, and one that has been adopted in many other previous analyses of non-standard dark energy models \cite{Maartens:astro-ph/0603353,Davis:astro-ph/0701510,2007ApJ...664..633W,2007PhRvD..76b3508G}.

Indeed, the physical motivation for adopting this approximation lies in the fact that both observables mostly depend only on the expansion rate of the universe. The CMB peak position is defined by a standard ruler (the sound horizon at recombination) which is very weakly affected by the dark energy component (firstly, because dark energy has essentially no influence before recombination, secondly, because the sound speed which enters into consideration here is that of the baryon-photon fluid, regardless of the other components). This standard ruler is then seen under different angles depending on the angular diameter distance at recombination. This, in turn, depends only weakly on the equation of state of dark energy, and more strongly on its density: in our model, at least for the values of the $\alpha$ parameters considered here, this follows very closely that of $\Lambda$CDM, hence it is reasonable to believe that the approximation does not introduce significant errors. The same considerations apply to the A parameter.  On this basis, we are then confident that the errors we are making by extending the $\Lambda$CDM approximation to our model lie well within the present accuracy of our analysis.

\section{Results} \label{results}
\begin{figure}[ht!]
   \includegraphics[width=0.9\columnwidth]{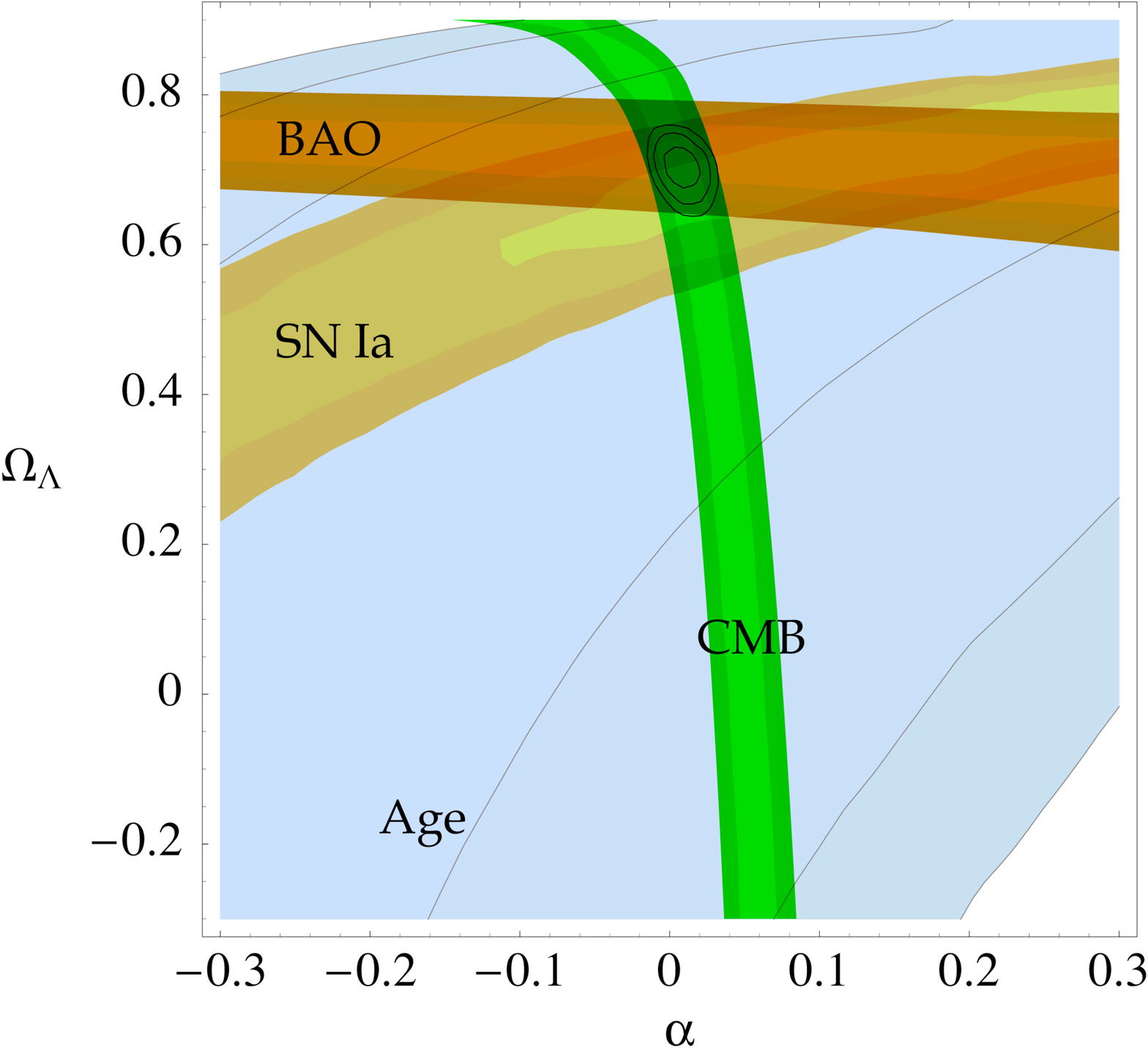}
    \caption{UDM likelihood contours at $68\%$, $95\%$ and $99\%$ C.L.\ (shaded regions), in the $\alpha$--$\Omega_\Lambda$ plane, obtained from the age of the universe, type Ia supernovae (SN Ia), the location of cosmic microwave background acoustic peaks (CMB), and baryon acoustic oscillations (BAO). Also shown (continuous curves) are the likelihood contours obtained from combining the datasets.\label{fig:likeall}}
\end{figure}
\begin{figure}[ht!]
   \includegraphics[width=0.9\columnwidth]{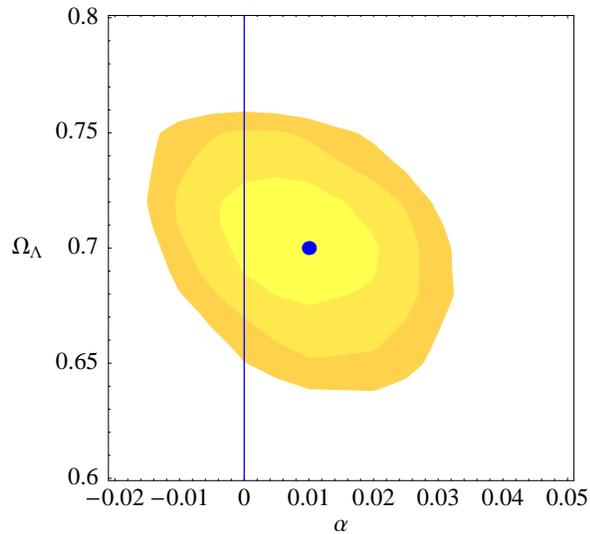}
    \caption{Likelihood contours at $68\%$, $95\%$ and $99\%$ C.L.\  in the $\alpha$---$\Omega_\Lambda$ plane, obtained from the combined datasets. The dot represents the combination of parameters which best fits our data. The vertical line locates the models describing  flat $\Lambda$CDM models. \label{fig:combined}}
\end{figure}
\begin{figure}[ht!]
   \includegraphics[width=0.9\columnwidth]{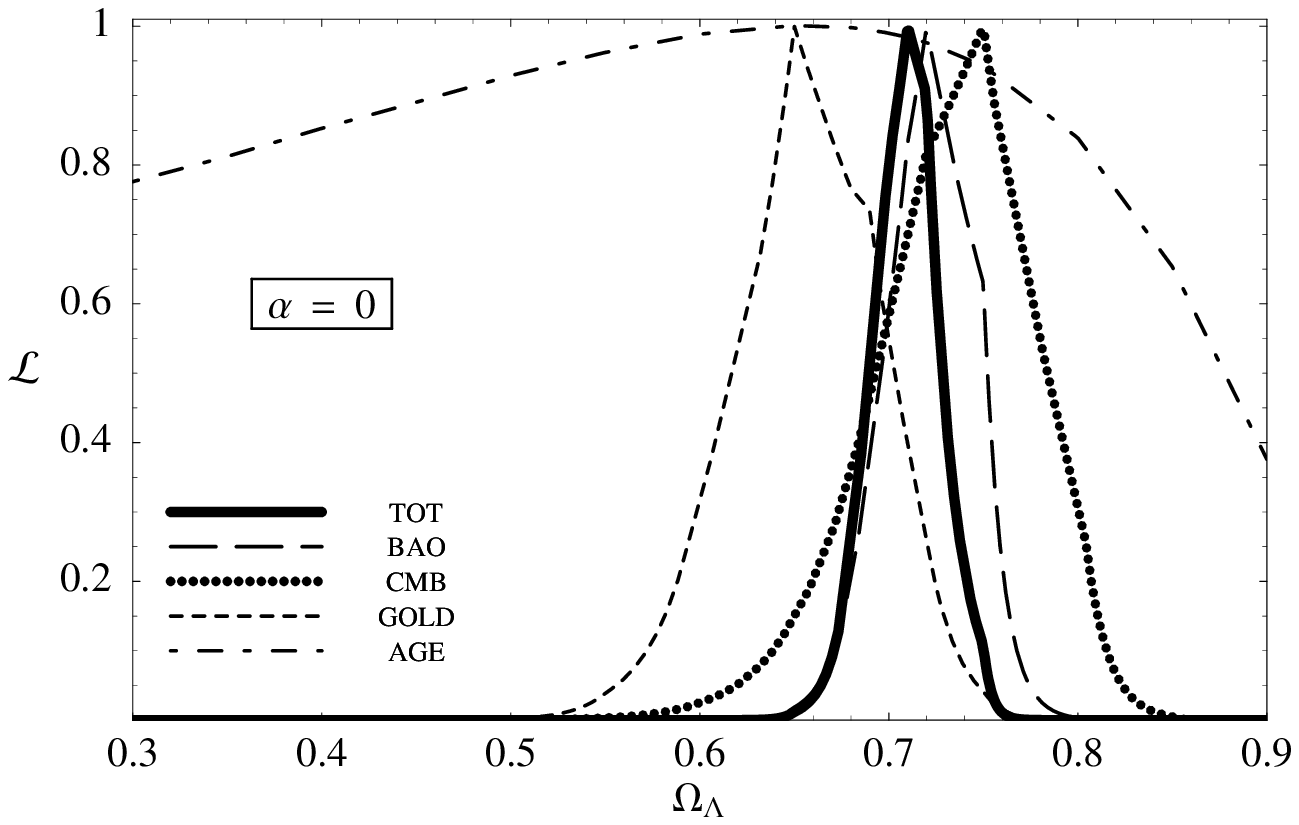}
   \includegraphics[width=0.9\columnwidth]{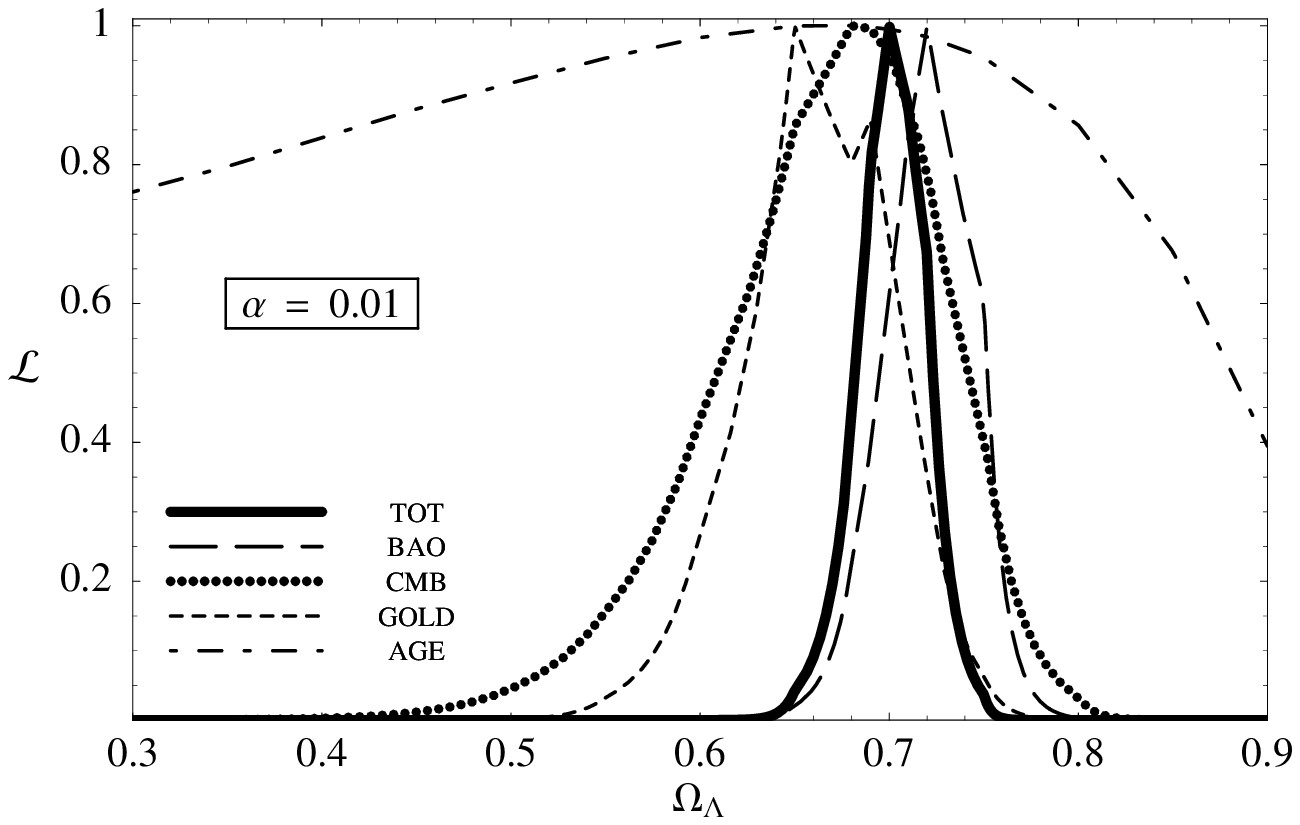}
    \caption{Marginalized likelihood functions for $\Omega_\Lambda$ derived from each dataset and from their combination. Top panel: the flat $\Lambda$CDM model, obtained when  $\alpha=0$ is assumed as a prior.  Bottom panel: the likelihood obtained by assuming the best fit value $\alpha=0.01$ for the $\alpha$ parameter. \label{fig:1dlikelihoods}}
\end{figure}

Fig.~\ref{fig:likeall} shows the confidence levels derived from the different cosmological observations used in our analysis, together with the joint confidence levels from the combined datasets.  The combined levels are also shown separately in more detail in Fig.~\ref{fig:combined}. A fairly narrow area of the parameter space is identified by our analysis, encompassing models which are close to the standard cosmological constant case with $\Omega_\Lambda\approx 0.7$. However, there is a preference for values of $\alpha$ larger than zero. Our best fit is $\Omega_\Lambda=0.70\pm 0.04$ and $\alpha=0.01\pm 0.02$, both at 95\% C.L.; for these values, we find $\chi^2=157.64$ with 185 data points. The marginalized likelihood functions for the $\Omega_\Lambda$ parameter are shown in Fig.\ \ref{fig:1dlikelihoods}. As said  above, when $\alpha=0$ is assumed as a prior, our model describes a flat $\Lambda$CDM model. In this case, we find a result which is consistent with previous analysis: $\Omega_\Lambda=0.71 \pm 0.04$ at  95\% C.L., with $\chi^2=158.83$.  We note that, for $\alpha=0$, the SNe data seem to prefer a slightly smaller value of $\Omega_\Lambda$ (see Fig. \ref{fig:1dlikelihoods}, top panel). This agrees with similar results (see e.g.\ \cite{Nesseris:astro-ph/0612653}, which also suggest that this might be due to some systematic effect in the latest SNe data).

In order to assess the statistical significance of our findings, we compare the $\Lambda\alpha$DM model with the   flat  $\Lambda$CDM  using the Akaike and the Bayesian Information Criteria (AIC and BIC, \cite{Liddle:astro-ph/0401198}), as well as  the Bayesian evidence $E$  \cite{Liddle:astro-ph/0608184} (see also \cite{2007MNRAS.378...72T,2006PhRvD..74b3503K}). 
In essence, all these  can be used as  model selection criteria, i.e.\ they estimate how much adding a new parameter to the cosmological model (in our case, the $\alpha$ parameter) is justified by the increased goodness of fit (cf. also \cite{Davis:astro-ph/0701510}). BIC improves on AIC taking also into account the number of data points. BIC  should be a crude approximation to  $E$, where the latter  is a further refinement as it averages the likelihood of the model in the prior: 
 $E\equiv\int{\cal L}(p)P(p)dp$ where $p$ are the free parameters of the model and $P$ is the prior distribution of the parameters. $E$ is precisely the likelihood of the model given the data, according to Bayes theorem, and as such gives positive weight to a ``good on average" fit over a larger volume in parameter space. The model comparison is achieved through computing  the quantity $\Delta\ln E=\ln E_{\Lambda 0}-\ln E_{\Lambda\alpha}$. We find $\Delta\ln E=3.9$, which has to be regarded as a rather strong evidence in favor of the flat $\Lambda$CDM model, given the present data \cite{Liddle:astro-ph/0608184}. Using AIC and BIC gives  milder  results (evidence but not strong evidence, cf.\  \cite{Liddle:astro-ph/0401198}), summarised in Table \ref{table1}.
 
\begin{figure}[ht!]
   \includegraphics[width=0.9\columnwidth]{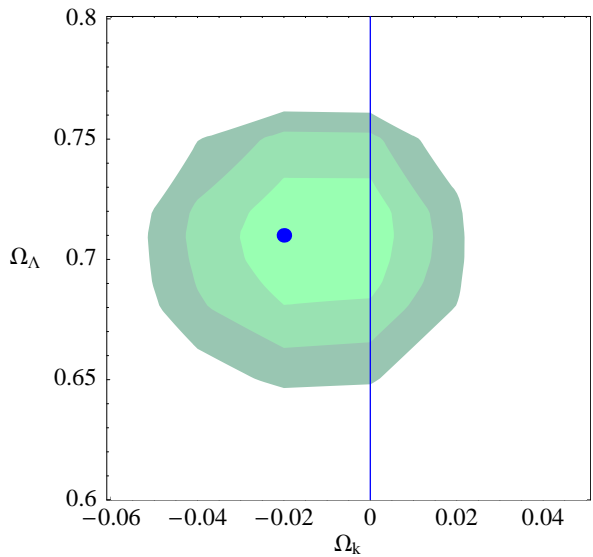}
    \caption{$\Lambda$CDM models: likelihood contours at $68\%$, $95\%$ and $99\%$ C.L.\  in the $\Omega_K$---$\Omega_\Lambda$ plane for the combined dataset. The dot represents the combination of parameters which best fits our data. The vertical line locates the flat ($\Omega_K=0$) models. \label{fig:lambdak}}
\end{figure}

  However, our analysis shows rather   clearly that the best fit model in our class  is not a flat $\Lambda$CDM and, as said in the introduction, recent results favor a slightly closed universe within the $\Lambda$CDM models \cite{Spergel:astro-ph/0603449}.  
Then, a  question worth investigating is which model  performs better in terms of AIC, BIC and  $E$ if we add an extra parameter to the flat $\Lambda$CDM,  either by allowing for curvature in the $\Lambda$CDM scenario, or through our $\Lambda\alpha$DM models, i.e.\  adding  $\Omega_K$ or $\alpha$, respectively.
We find that for non-flat $\Lambda$CDM models
the best fit is achieved for $\Omega_\Lambda=0.71\pm 0.04$ and $\Omega_K=-0.02^{+0.01}_{-0.02}$,   see  Fig.~\ref{fig:lambdak}, and again there is evidence in favor of the  flat $\Lambda$CDM, see Table \ref{table1}.  
A direct comparison of  the {\it not nested} \cite{Liddle:astro-ph/0401198} non-flat  $\Lambda$CDM and the $\Lambda\alpha$DM models is possible in terms of AIC and BIC: in terms of the latter, we see  from Table \ref{table1} that there is positive evidence in favor of  the   $\Lambda\alpha$DM model. An indirect comparison of the $\Delta \ln E$'s suggests however  that the two models are statistically almost equivalent.

\begin{table}[htdp]
\begin{center}\begin{tabular}{|c|c|c|c|c|}\hline Model & $\chi^2$ & $\Delta$AIC & $\Delta$BIC & $\Delta \ln E$ \\\hline flat $\Lambda$CDM & 158.83 & 0 & 0 & 0 \\\hline flat $\Lambda\alpha$DM & 157.64 & 0.8 & 4 & 3.9 \\\hline curved $\Lambda$CDM & 161.53 & 4.7 & 7.9 & 3.4 \\\hline \end{tabular} \caption{Model comparison with information criteria and Bayesian evidence: the $\Delta$'s compare the  flat $\Lambda\alpha$DM and the non-flat $\Lambda$CDM against the standard flat $\Lambda$CDM model. \label{table1}}
\end{center}
\end{table}

\section{Conclusions}
Today's standard model for the homogeneous isotropic  universe is the 1-parameter flat  $\Lambda$CDM, based on GR as the theory of gravity, a flat spatial geometry as it follows from the inflationary paradigm,  cold dark matter and an {\it ad hoc} cosmological constant $\Lambda$  as the only two relevant constituents apart from baryons and radiation.
However, the now well established  accelerated expansion of the universe calls for an explanation beyond the {\it ad hoc}  $\Lambda$ term in Einstein equations. 
In the game of models testing and selection, in absence of really compelling models from fundamental physics,  that of simplicity of models - the Occam's razor - remains a critical ingredient. Quantitatively, this can be implemented using Bayesian methods in the statistical analysis of data, thereby 
measuring the worth of models by favoring those that give a good fit with fewer parameters. On the other hand,  it does makes sense in our view to consider models built on some physical basis, and compare models with equal number of parameters but different physical motivations, in an attempt to establish which model performs better.

Here we have assumed a single dark component with barotropic EoS $P_X=P_X(\rho_X)$ to make up for a total $\Omega=1$ (i.e. a  flat model) and the acceleration of the universe. Assuming a constant speed of sound $c_s^2=\alpha$ the 2-parameters affine EoS (\ref{EOSaffine}) then follows: we then dub our model
 as $\Lambda\alpha$DM, as it admits an {\it effective} cosmological constant and thus  it is formally equivalent to having a $\Lambda$ term and an evolving part characterized by $\alpha$. 
For $\alpha=0$, our UDM  is formally equivalent to $\Lambda$CDM  for today's density value $\rho_{Xo}>\rho_\Lambda$ in (\ref{rhodia}). On the other hand, for any $\alpha>-1$ and for  $\rho_{Xo}<\rho_\Lambda$ our UDM  would be phantom, with no ``big rip" \cite{Ananda:astro-ph/0512224}.
 
Simple parameterizations of the EoS may lead to misleading results \cite{Bassett:astro-ph/0407364,Riess:astro-ph/0611572}. 
 Comparing 2-parameter   EoS's,
 in our  view  barotropic forms\footnote{See e.g.\ \cite{Davis:astro-ph/0701510} for a recent observational test of the Chaplygin gas and other examples.}  like  (\ref{EOSaffine})  have at least the advantage over the common scale factor or redshift parameterizations  of relating directly   $P_X$ and $\rho_X$. This seems more physical for an EoS: one wants the parameterization to express an intrinsic and realistic relation between the fluid or field \cite{Crittenden:astro-ph/0702003} variables, disentangling it from   universe evolution.  
In addition, an EoS of the form $P_X=P_X(\rho_X)$ has a validity that goes beyond that of the homogeneous isotropic background. It can indeed be easily carried forward to study inhomogeneities:  at perturbative order one can use it assuming adiabatic perturbations with $\delta p= c_s^2\delta \rho$, representing either a barotropic fluid or a k-essence field \cite{Quercellini:2007ht}, or else to formally describe a scalar field with unitary effective speed of sound, $c_{eff}^2=1$ \cite{BED91,Hu:astro-ph/9801234,Quercellini:2007ht}. We are addressing the growth of perturbations and their CMB signatures for all these cases in a forthcoming paper \cite{next}.

Constraining the $\Lambda\alpha$DM model with various observables we find that it fits the data  well, better than   $\Lambda$CDM, with best fit values $\alpha=0.01 \pm 0.02$ and $\Omega_\Lambda=0.70 \pm 0.04$. From Eq.\ (\ref{eq:w0})  we then get $w_{X0}\simeq -0.7$ at leading order: indeed we find that  values of $\alpha$ and $\Omega_\Lambda$ corresponding to a phantom dark component are ruled out  well beyond $3\sigma$'s. Our results for model comparison are summarized in Table \ref{table1}. The flat $\Lambda$CDM model is still preferred, with a ``rather strong" evidence \cite{Liddle:astro-ph/0608184}. Comparing our model with a non-flat $\Lambda$CDM shows however that the two are, all in all, almost statistically equivalent.
Then, one would advocate our model, which is assumed to be flat, on the basis of an inflation motivated  theoretical bias.

Beyond this factual analysis, we may as well look at Fig.\ \ref{fig:likeall} from a different perspective. it is indeed rather clear that BAO and SNe data alone prefer a larger $c_s^2=\alpha$ value than our best fit value. Indeed, we find  that the BAO+SNe best fit is $\alpha\approx0.1$. On the other hand, it is also clear from Fig.\ \ref{fig:likeall} that the CMB strongly constrains $\alpha$, and forces it to take a very small value (cf.\ \cite{Muller:astro-ph/0410621,next}). This seems to suggest that high redshift data (CMB) favor $c_s^2\simeq 0$, while low and intermediate redshift data (BAO and SNe) allow for a larger $c_s^2$, i.e.\ that the speed of sound of a UDM component should be variable. This is indeed in line with the findings of an analysis of models such as the generalized Chaplygin gas, which  shows indeed a very good fit to data \cite{Davis:astro-ph/0701510}. As we have shown here, the standard  $\Lambda$CDM can be seen as a UDM with vanishing speed of sound at all redshifts, $c_s^2=\alpha=0$.  Thus, If proved by future observations, a non vanishing speed of sound at low and intermediate redshift for the whole dark component in the Friedmann equation (\ref{einstein1}) would provide a sort of observational ``no-go theorem" for the standard $\Lambda$CDM in GR.

In this paper, in order to consider a minimalist 2-parameter model, we have assumed a single UDM  component with constant speed of sound $c_s^2=\alpha$  in addition to baryons and radiation. It could well be  that the absence of collisionless CDM  is going to spoil structure formation in this model, or else, that our model is subject to much stronger constraints, forcing it to coincide with $\Lambda$CDM in practice, when a full CMB analysis is carried out (cf.\ \cite{Muller:astro-ph/0410621,next}).
On the other hand the simple  $\Lambda\alpha$DM component, derived assuming a constant speed of sound,  may turn out to be a useful and physically motivated model for  dark energy  (cf. \cite{Babichev:astro-ph/0407190,Holman:astro-ph/0408102}), additional to CDM. 
In any case, it can be shown \cite{Quercellini:2007ht} that the affine EoS
(\ref{EOSaffine}) corresponds to an exact solution for a quintessence scalar field with appropriate potential, as well as to the general dynamics of a k-essence field.  The quintessence field in particular  is going to have different
perturbations than the corresponding fluid (see e.g.\ \cite{BED91,Hu:astro-ph/9801234}), which then need to be separately analized.
All these  problems can only be settled with  future work. 

\begin{acknowledgments}
We thank members of ICG, Portsmouth, for useful discussions, in particular Robert Crittenden. MB  thanks MIUR for a "Rientro dei Cervelli" grant and the Galileo Galilei Institute for Theoretical Physics (Florence) for hospitality while part of this work was carried out, and the INFN for partial support during the visit.
\end{acknowledgments}

\bibliography{BBQ}

\end{document}